\documentstyle[12pt]{article}
\input epsf
 
\begin{document}

\begin{titlepage}
\null\vspace{-62pt}
 
\pagestyle{empty}
\begin{center}
\rightline{CCNY-HEP-00/5}
\rightline{hep-ph/0009225}
 
\vspace{1.0truein}
{\Large\bf Non-topological solitons as nucleation sites
            for cosmological phase transitions  }
 
\vspace{1in}
D. METAXAS \\
\vskip .4in
{\it Physics Department\\
City College of the City University of New York\\
New York, NY  10031\\
metaxas@theory.sci.ccny.cuny.edu}\\
 
\vspace{0.5in}
 
\vspace{.5in}
\centerline{\bf Abstract}
 
\baselineskip 18pt
\end{center}

I consider quantum field theories that admit
charged non-topological solitons of the Q-ball
type, and use the fact that in a first-order
cosmological phase transition, below the critical temperature,
there is a value of the soliton charge above which the soliton
becomes unstable and expands,
converting space to the true vacuum,
much like a critical bubble in the case
of ordinary tunneling.
Using a simple model for the production rate of Q-balls
through charge accretion
during a random walk out of equilibrium,
I calculate the probability
for the formation of  critical charge
solitons 
and estimate the amount of supercooling
needed for the phase transition
to be completed.

\end{titlepage}

\newpage
\pagestyle{plain}
\setcounter{page}{1}
\newpage

\section{Introduction}
 
A common feature
of various cosmological models with a first-order
phase transition is the generation of
excessive entropy. This happens
Because the probability of tunneling
from the false to the true vacuum
is negligible when the supercooling
parameter $\eta=(T_c -T)/T_c$ is
very small.
The question arises then if there exists
some other mechanism that can
facilitate the completion of the transition
more rapidly, without the problems
of excessive entropy production.
 
In other cases of interest for cosmology
a relatively large amount
of supercooling is actually needed in order
to have out-of-equilibrium processes
such as the generation of baryon asymmetry.
A detailed picture of the phase transition
is then necessary.

Here I consider a different mechanism
for a 
phase transition that is possible in theories
that admit non-topological solitons of the
Q-ball type, and show that the rate
of their nucleation can be large enough
for the transition to proceed with much
less supercooling than in other cases. 
 
There is  a  large amount of  literature on
non-topological solitons  (see
\cite{Lee-Pang}
for a review).
Here I will be interested in the case
of Q-balls that are formed  in the context
of a first-order cosmological phase transition.  
 
The production of non-topological solitons
in a cosmological phase transition has
been studied before \cite{Frieman, Kolb2},
and I will now describe briefly some
previous results, focusing in the case
of a first-order transition.
 
Assume a potential $U(\phi, T)$ with
a symmetry preserving minimum at $\phi=0$
and a symmetry breaking minimum at $\phi=\phi_+$,
separated by a barrier with a local maximum $U_M$
at $\phi=\phi_M$, 
and consider the evolution of the vacuum
during the phase transition.
At high enough temperature $T$ the universe
is in the symmetric vacuum. At temperatures
close to the critical temperature $T_c$,
thermal fluctuations of the field $\phi$
become important. The fluctuating regions
typically have volume  
$V_{\xi}=\frac{4\pi}{3}\xi^3$, where
$\xi\sim\frac{1}{m_{\phi}(T)}$ is the 
correlation length.
The relative probability of finding
a certain configuration of  the
field $\phi$ in a  correlation volume
is $P\sim\exp(- F/T)$, 
where $F$ is the free energy
of the given configuration.
 
For high enough temperatures
thermal  fluctuations are strong
enough to flip regions of the 
correlation volume from the
symmetric to  the asymmetric minimum
and vice-versa.
When the temperature falls  below 
a certain temperature called the
Ginsburg temperature $T_G$, thermal
fluctuations are not strong enough 
to mediate such transitions.
Since the number density of
the $\phi$ particles   is
$n_{\phi}\sim T^3$, we will  have
$N \sim T_G^3 V_{\xi}$ charged particles
in a correlation volume,
and we expect a net charge
fluctuation $Q_{b}\sim\sqrt{N}$
in each bubble of the asymmetric phase
(I assume a case with zero
initial charge asymmetry).
 
Usually, in theories that admit
Q-balls, there is a
constraint of a minimum charge $Q_{min}$ that they
have to exceed for reasons of stability.
If $Q_b >Q_{min}$ then we expect
a Q-ball to be formed in the bubble of
the asymmetric phase. 
 
In the case of a first-order
phase transition, when the symmetric
vacuum becomes metastable, the
potential energy density in the interior
of the Q-balls becomes negative 
with respect to the potential energy density
of the symmetric vacuum.
Then if a Q-ball has  large enough charge
it becomes unstable and expands, converting
space  to  the true vacuum, much  like
a critical bubble in the case of ordinary
tunneling \cite{Spector, Kusenko}.  
 
In previous discussions of Q-ball
production in the early universe,
the constraint of the minimum charge,
was  quite restricting for the occurence
of the phase transition, especially since
most of the estimates were made using
the thin-wall  approximation for the solitons,
which leads to a relatively large minimum
charge \cite{Ellis}.
 
It was recently shown that the thin-wall
assumption is not correct,
and that some theories may 
admit Q-balls without the
restriction of a minimum charge,
or with a minimum charge
of order one. 
Then, 
charge accretion may lead to a growth
in the number density of the critical charge solitons
below some temperature, and thus mediate
the phase transition 
\cite{Kusenko}.
An estimation of the nucleation rate
of these solitons is then important
for the description of the phase transition.

In the case where the solitons that are initially
produced survive long enough to reach thermal 
equilibrium, and if  a set  of reactions that produce
critical solitons remain in thermal equilibrium,
then critical solitons are copiously produced. 
The conditions for this to happen are quite
model-dependent \cite{Kusenko, Kolb2}. Here  
I  will deal with the case of critical charge
Q-balls appearing
before  thermal equilibrium, and show that
 the rate of their
production by charge accretion can  be
large enough for the phase transition
to proceed.   
This case may be relevant also when    
we have soliton destruction or
evaporation before they reach thermal equilibrium.  

I will begin by reviewing various approximations that
already exist for the description of Q-balls in 
various ranges of the charge $Q$.
Then I will focus on the case of a first-order 
phase transition in theories
that admit non-topological solitons
 and calculate the rate
of Q-ball production using results
from subcritical bubble fluctuations
and charge fluctuations.
At temperatures near the critical temperature
thermal fluctuations may flip regions from
the symmetric to the asymmetric vacuum.
If these regions also contain a charge fluctuation $Q$,
and have energy less than $Q$ free particles,
then, instead of decaying into free particles, they
will evolve into the soliton configurations.
The evolution of these bubbles into solitons
is affected by their interaction with the thermal
environment. An absorption of a positive or negative
charge can be simulated by a random walk with a
$+$ or $-$ step. The difference from
an unrestricted random walk is the
existence of a minimum charge below which they 
disappear. This is the case of a random walk
with an absorbing boundary. Using this model I will
calculate the final soliton production rate.
Once they are produced these solitons keep interacting
with the medium, and their evolution can still
be simulated by a random walk. Below the critical temperature
there is also a critical soliton charge, above which the
solitons become unstable. This is
the case of a random walk with two
absorbing boundaries.
Using this model I will calculate the production rate
of critical solitons. The critical solitons expand converting space
to the true vacuum and mediating the phase transition.
I will then estimate the amount of supercooling needed
for the nucleation of a cosmological phase transition.
The main result of this work is that, for
generic values of the model's parameters, 
the phase transition can proceed much faster, with
a much smaller amount of supercooling 
than other models (that do not admit nontopological
solitons).

This result may be important for 
cosmological phase transitions in theories
that admit non-topological solitons,
such as the Minimal Supersymmetric 
Standard Model (MSSM) \cite{Kusenko},
\cite{MSSM}, or phenomenological 
theories of hadronic matter \cite{Lee-Pang}.
 
In Sec.~2 I  describe 
some  approximations that already
exist for the description of Q-balls
in various regimes  
\cite{Coleman,
Kusenko,
Spector}.
In Sec.~3   I will model the
production of Q-balls during the phase
transition, using results
from subcritical bubble fluctuations
\cite{Kolb1}
and charge fluctuations
\cite{Kolb2}.
I will then calculate the probability for
the formation of critical
charge Q-balls through charge accretion
and in Sec.~4 I will estimate the 
amount of supercooling needed
for the completion of the phase transition.

\section{Non-topological solitons in various approximations}

Consider a quantum field theory 
of a complex scalar field $\phi$
with Lagrangian
\begin{equation}
L=\frac{1}{2}(\partial_{\mu}\phi)^2 - U(\phi)
\end{equation}
that is admits a  global
$U(1)$ symmetry
and a   
conserved charge
\begin{equation}
\phi\rightarrow e^{i\theta} \phi \,\, ,  
\,\,\,\,\,\,\,\,\,
Q=\frac{1}{2i}\int d^3 x
\,     \phi^{*} \stackrel{\leftrightarrow}{\partial_t}\phi
\end{equation}
Minimizing the   
energy
in the sector of field configurations that
have a given charge $Q$ 
yields the three-dimensional  bounce equation
\begin{equation}
\nabla^2\,\phi = U_{\omega}^{'}(\phi)
\end{equation}
where 
\begin{equation}
U_{\omega}(\phi)=
 U(\phi)-\frac{1}{2}\omega^2\phi^2  \,\, ,
\,\,\,\,\,\,\,\,\,\phi(x,t)=e^{i\omega t}\phi(x)
\label{charge}
\end{equation} 
This bounce equation has a solution
when $U_{\omega}(\phi)$ has a nonzero
global minimum in addition to
the local minimum at the origin.
In order for this to happen,
$\omega$ must be less than $M$,
where $M^2 = U''(0) $, and greater
than some value $\omega_0$, for which
$U_{\omega}(\phi)$
has two degenerate minima.
Minimizing  
$E_{\omega}$ with respect to $\omega$ we get  
the soliton energy $E_Q$.
 
I will be interested in the case
of nontopological solitons that
appear near a first-order phase
transition for which I will consider
a generic effective potential
depending on the temperature $T$:   
\begin{equation}
U(\phi,T)=\frac{1}{2}M^2(T) \phi^2
          -A(T)\phi^3 + \lambda \phi^4
\label{ftpot}
\end{equation}
where
\begin{equation}
M^2(T)=(\alpha T^2 -m^2) 
\end{equation}
for which
\begin{equation}
U_{\omega}(\phi)=\frac{1}{2} (M^2 -\omega^2)\phi^2 -A \phi^3
        + \lambda \phi^4 \,\,.
\end{equation}
The cubic term is understood  as an  effective
term of the form $(\phi^* \phi)^{3/2}$
that still has a global $U(1)$ symmetry,
and $\alpha$ is the square of the gauge
coupling constant, assumed to
be greater than $\lambda$.
The coefficient $A$ will also depend on the temperature,
but I will neglect this dependence near the critical temperature.  
 
In the limit of large $Q$,  which
is equivalent to 
$\omega\rightarrow\omega_0 +0$,
the form of the Q-ball solution
simplifies considerably \cite{Coleman}
since       
$U_{\omega}(\phi)$ is such that
the thin wall approximation 
is valid for the
bounce solution.
 
In the opposite limit
of $\omega\rightarrow M$ we see
that the barrier in the function $U_{\omega}(\phi)$,
which  provides the bounce solution,
becomes very shallow until it disappears.
In this approximation, neglecting     
 the $\lambda \phi^4$ term, 
and after rescaling  
\cite{Kusenko}  
\begin{equation}
\phi = \psi\frac{M^2 -\omega^2}{A} \,\,\,\,\,,\,\,\,
 x=\xi\frac{1}{\sqrt{M^2 -\omega^2}}
\end{equation}
one gets a minimum  
\begin{equation}
E_Q = QM(1-\frac{\epsilon^2}{6})
\label{EQ}
\end{equation}
if 
\begin{equation}
\epsilon\equiv\frac{Q A^2}{3 S_{\psi} M^2} < \frac{1}{2}
\label{defepsilon}
\end{equation}
where $S_{\psi}=4.85$ is the bounce action in
the dimensionless potential
$\frac{1}{2}\psi^2 -\psi^3$.
The bounce radius in $\xi$ is of order one
so the soliton radius is 
\begin{equation}
R_s \sim \frac{1}{\sqrt{M^2 -\omega^2}} \sim \frac{1}{\epsilon M}
\label{pl1}
\end{equation}

Now we consider the case
where the previous
potential  is at, or slightly below, the critical
temperature where it has two degenerate
minima at $0$ and $\phi_+$ :
\begin{equation}
U(\phi)=\lambda\phi^2(\phi -\phi_+)^2
\end{equation}
We consider large charge Q-balls in 
the thin wall 
approximation, but now the surface terms
are important since the two minima are degenerate,
and $\omega_0 =0$ since the original potential
has already a bounce solution.
 
The  Q-ball energy in the thin wall limit is 
\begin{equation}
E(Q) = 
     \frac{Q^2}{2\phi^2 V} +
         4\pi R^2 S_1 - \varepsilon V
\end{equation} 
where
\begin{equation}
\varepsilon = |U(\phi_+)|
\,\,\,\,\,\,\,\,\,   
S_1 
     =\int_0^{\phi_+} d\phi \sqrt{2U(\phi)}
\end{equation}
An approximate expression for the radius if 
we neglect the potential term is 
\begin{equation}
R^5 = \frac{9 Q^2} 
           {(8\pi)^2 \phi^2  S_1}
\label{pl2}
\end{equation}

As the temperature drops further from 
the critical temperature, the negative
potential volume term becomes important,
and there exists a value of $\varepsilon$
such that the Q-ball energy does not have
a minimum, and is unstable against expansion.
Equivalently, for a value of $\varepsilon$,
there exists a critical value of the charge $Q$,
such that for $Q>Q_c$ the Q-ball is unstable 
and expands converting space to the true vacuum.
These critical solitons have a charge $Q_c$
and a radius $R_c$ which are
obtained from
the simultaneous solution
of the equations
$
\frac{dE}{dR} = \frac{d^2 E}{dR^2} = 0
$ to be 
\cite{Kusenko}  
\begin{equation}
Q_c = \frac{400\,\pi\sqrt{70}}{81} 
            \frac{\phi_+\,S_1^3}
                 {\varepsilon^{5/2}}
\label{qc}
\end{equation}
and
\begin{equation}
R_c =  \frac{10}{3} \left( \frac{7}{8} \right)^{1/5}
         \frac{S_1}{\varepsilon}
\label{rc}
\end{equation}

\section{ Nucleation of critical Q-balls}

In this Section I will show that critical
charge Q-balls can be produced through charge
accretion from smaller charge Q-balls
during a random walk out of equilibrium,
and estimate the rate 
for this process. 
 
I will use the generic finite temperature
potential of the form
(\ref{ftpot}).
At the critical temperature $T_c$ we have
\[
V=\frac{1}{2}M^2\phi^2 -A\phi^3 + \lambda \phi^4
 =\lambda \phi^2(\phi -\phi_+)^2
\]
where $\phi_+$ is the second degenerate minimum,
and accordingly the parameters
of the potential satisfy the relations
\begin{equation}
M^2 = 2\lambda\phi_+^2  \,\,,\,\,
A=2\lambda\phi_+        \,\,,\,\,
M^2 = A^2 /2\lambda
\label{relations}
\end{equation}
The  surface term is   
\begin{equation}  
S_1 = 
   \int_0^{\phi_+} \,d\phi\,  \sqrt{2V(\phi)}\,=  
 \,   \frac{\sqrt{2\lambda}}{6}\, \phi_+^3
\label{surfenergy}
\end{equation}
and  at  a temperature $T=T_c - \delta T$
slightly below the critical temperature
the potential energy difference of the 
two minima is        
\begin{equation}
\varepsilon =  |V(\phi_+(T))| =
\alpha T \delta T \phi_+^2  =
\alpha T^2 \phi_+^2 \eta
\label{fvenergy}
\end{equation}
where $\eta = \delta T /T $
is the supercooling
parameter.

I will estimate the production rate
of Q-balls  at the critical temperature
using results of subcritical  bubble calculations
\cite{Kolb1}
and charge fluctuations
\cite{Kolb2}.
Subcritical bubbles are bubbles of
the symmetry breaking minimum
that have smaller than the critical radius.
In order to get an approximation
for the production rate of
subcritical bubbles
we will use the ansatz
for Gaussian-shaped bubbles
of radius $R$
\begin{equation}
\phi_b (r) = \phi_+ e^{-r^2 /R^2}
\label{scbubble}
\end{equation}
and  write an approximate expression
for their thermal production rate
\begin{equation}
\Gamma'(R) = a T^4 e^{-F(\phi_b)/T}
\label{appnucl}
\end{equation}
where $a$ is an unknown numerical factor of order one
and $F(\phi_b)$ is the bubble free energy
\begin{equation}
F(\phi_b)= \int \, 4\pi r^2 \, dr
           \left( \frac{1}{2} \left(\frac{d\phi}{dr}\right)^2
                  + V(\phi,T) \right)
         = \gamma\, R + \delta \, R^3 \,\,.
\label{Fb}
\end{equation}
For a bubble of the form (\ref{scbubble})
and the potential (\ref{ftpot})
at the critical temperature,
it turns out that the potential term
will give the largest 
contribution to the production rate
below,   
with a coefficient \cite{Kolb1}
\[   \delta = \pi^{3/2} \phi_+^2
     \left( \frac{\sqrt{2}}{8} M^2 
           -\frac{\sqrt{3}}{9} A \phi_+
           + \frac{\lambda}{8} \phi_+^2
        \right)
 \,=\, c_1 \, \phi_+^2 \, M^2  
\]
where $c_1 = 0.26 $.
 
Now, in a typical bubble of radius $R$ there are
\[
  N \approx  c_2 \, T^3 R^3
\]
$\phi$ and $\bar{\phi}$ particles,
where $c_2 =\frac{4\pi\zeta (3)}{3\pi^2}\,= 0.51 $.
The probability of a fluctuation
of total charge $Q$ in this number is
\cite{Kolb2} 
\[
   P(Q\,;\,N) =
   \frac{e^{-Q^2 /N}}{\sqrt{N}}
\]
Then, changing variables from $R$ to $N$,
the total production rate of bubbles
of charge $Q$ is
\begin{eqnarray}
\Gamma'(Q)&=& \int \Gamma(N) P(Q\,;\,N) \, dN
          =  \nonumber \\
    &=&  3\sqrt{c_2} \, a \,T^{11/2}
         \int e^{-\frac{c_1 \phi^2 M^2 R^3}{T} 
                 -\frac{Q^2}{c_2 T^3 R^3} }  \,\,
             \sqrt{R} \, dR 
\end{eqnarray}
where from now on we write
$\phi$ for the symmetry breaking minimum
$\phi_+$.
We can evaluate this integral using the saddle
point method. The exponent has a minimum
at a radius
\begin{equation}
R_b^3 = \frac{Q}{\sqrt{c_1 c_2}\phi M T} \,\,.
\label{sbradius}
\end{equation}
Absorbing some numerical constants in the
prefactor $a$ of order one,
and denoting a new numerical constant  
$c=2\sqrt{c_1}/\sqrt{c_2}= 1.42  $
we get the final  result
\begin{equation}
\Gamma'(Q) = a \, \frac{T^{13/2}}
                    {(\phi M)^{5/4} \, Q^{1/4}} \,
              e^{\textstyle -c\,\frac{Q\,\phi\,M}{T^2}}
\,\,.
\label{nuclrate}
\end{equation}
 
The above expression for the
nucleation rate  $\Gamma'(Q)$ of bubbles of the true
vacuum with net charge $Q$ is still not 
quite the same as the nucleation rate
of Q-balls.
First, the bubble energy should be smaller than
the corresponding energy of $Q$ free particles,
otherwise the bubble will decay into free particles
instead of evolving into a soliton.
From 
our previous expressions
we see that, for generic
values of our parameters, 
this is indeed the case,
and accordingly the charged bubbles,
during their subsequent interaction
with the particles of the heat bath,
will preferably  evolve to the
soliton configuration
instead of decaying into free particles.
Although the initial bubble energy is not
exactly the same as the corresponding 
soliton energy, the thermal fluctuations
from the surrounding heat bath 
would rapidly compensate for that
during the subsequent random walk.
It should be noted, however, that 
the fact that the bubble energy is less than
the corresponding energy of $Q$ free particles is
also a model-dependent condition and has to be checked
in different models.

Second, the Q-balls are
bounces in a modified potential, and
may be very different configurations from 
the true vacuum bubbles. 
If, for example, 
the bubble radius is much smaller
than the corresponding soliton radius   
then it will expand
until it becomes the lowest energy configuration
of a Q-ball, but in the process it will
meet a number of particles contained in this
volume, and if there is a net charge fluctuation
of negative charge $Q$, then the bubble
charge will disappear.
From the expressions  for the 
corresponding radii of  the subcritical
bubbles and the solitons,
(\ref{sbradius}) and (\ref{pl1}),
and using (\ref{defepsilon}) 
we get
for the ratio of the volumes
of the bubble and the soliton of charge Q:
\begin{equation}
\left(\frac{R_b}{R_s}\right)^3 \sim
\frac{Q}{\sqrt{c_1 c_2}\phi M T} \, \epsilon^3 M^3
= 40 \epsilon^4 \frac{M^4}{A^2 \phi T}
\end{equation}
From (\ref{relations}) we have:
\begin{equation}
M^2 \sim \alpha T^2 \sim \lambda  \phi^2\,,\,\,
A \sim \alpha \phi
\end{equation}
thus
\begin{equation}
\left(\frac{R_b}{R_s}\right)^3 \sim
40 \epsilon^4 \sqrt{\frac{\alpha}{\lambda}}
\end{equation}
with 
\begin{equation}
\epsilon \sim Q A^2 /M^2 \sim Q \lambda
\end{equation}
and for charges of  order  one we see that
$\epsilon$ is  much less than one, 
in which case we indeed have the situation  
mentioned above.
 
During the expansion of the bubble  to
a soliton, the bubble will encounter
a number of positive and negative
charges $N$ which are contained in the
volume of the soliton.
This process   
can be simulated 
by a random walk, with a $+$ ($-$) step if
a Q-ball absorbs a positive (negative)
charge.
The difference from an unrestricted random walk
is    the existence of a minimum charge below which
the Q-balls are unstable and disappear,
This is the situation in a random walk
with an  absorbing boundary,     
and the probability that 
an initial charge $Q'$ will evolve to 
a charge $Q$ is given by
\cite{Feller}:
\begin{equation}
P(Q' ,Q) = \frac{Q Q'}{N^{3/2}}\,
           e^{-\frac{Q^2 +Q'^2}{2 N}}
\label{fell1}
\end{equation}
where $N$ is the number of particles
contained in the soliton volume
(from now on we will use $Q$ and $Q'$
to denote the units of charge
{\it above} the minimum charge $Q_{min}$).

Hence, the final expression for the nucleation
rate of solitons of charge $Q$ is
\begin{equation}
\Gamma(Q)= \int_0^{\infty} \Gamma'(Q')
              P(Q', Q) dQ'
\end{equation}
Using the previous equations
(\ref{nuclrate}), (\ref{fell1}), (\ref{pl1}),
we get
the final result for the nucleation rate of 
solitons
of charge $Q$ 
\begin{equation}
\Gamma(Q)=\Gamma_m
  \frac{\lambda^{9/2} M^{5/2}}{T^{1/2} \phi^2}
 \, Q \, 
 e^{-\frac{2 \lambda^3 M^3}{T^3} \, Q^2}
\label{neweq}  
\end{equation}
where
\begin{equation}
    \Gamma_m = a \, \frac{T^{13/2}}
                      {(\phi M)^{5/4}} 
             e^{\textstyle -c\,\frac{Q_{min}\,\phi\,M}{T^2}}
    \label{gammam}
  \end{equation}
and again we absorbed numerical factors of order one
in the prefactor $a$.

Once the Q-balls are formed and start
interacting with the 
positive and negative charges 
in the heat bath,
their subsequent  evolution, before they
reach thermal equilibrium or disappear,
can be simulated again  
by a random walk, with a $+$ ($-$) step if
a Q-ball absorbs a positive (negative)
charge.
The difference from an unrestricted random walk
besides  the existence of a minimum charge below which
the Q-balls are unstable and disappear,
is  the existence of a critical charge
above which the Q-balls are unstable and
expand converting space into the true vacuum.
This is the situation in a random walk
with two absorbing boundaries.
 
We are interested in the probability for
a Q-ball, starting from an initial charge $Q_0$
which is arbitrary but small enough so
that the rate for its nucleation is
not negligible, to reach the critical
charge $Q_c$ before it reaches the minimum charge,
and  we also need an estimate of the
total time that is needed for this process.
 
Let $X(Q,N)$ be the probability that
a point at initial position $Q$
will have crossed $Q_c$ by $N$ steps
before crossing $0$. Let $p$ and $q$ be the
relative probabilities for up and down motion
in a single step, with $p+q=1$.
For sufficiently large number of steps
$X(Q,N)$ becomes independent of $N$.
In this case it is easy to calculate
this probability  $X(Q)$.
Assuming there is  no asymmetry
between positive and negative charges,
and $p=q=1/2$,
we get\footnote{Note that we have shifted
the charges by $Q_{min}$ so the exact
expression for (\ref{probability}) would
be  $X(Q)=\frac{Q}{Q_c -Q_{min}}$ but since
$Q_c \gg Q_{min}$ the above expression
(\ref{probability}) would suffice.} 
\cite{Feller}
\begin{equation}
X(Q)=\frac{Q}{Q_c}
\label{probability}
\end{equation}
and the nucleation rate for 
critical charge solitons is
\begin{equation}
\Gamma(Q_c) = \int \, dQ  \, X(Q) \Gamma(Q)
\label{ccc}      
\end{equation}

For applications to a realistic situation
we are interested in a different kind of 
random walk, one where the time step
is variable, depending on the value of
the Q-ball charge at any step, since
different  charges have different
cross  sections.
If $n(\phi)$ and $v$ are the density and
velocity of the particles in the thermal bath
and $\sigma (Q)$ is the cross section
for their absorption by a Q-ball of  charge $Q$,
then
the time step varies as
\begin{equation}
\Delta t \,\sim\,\frac{1}{n(\phi)\,\sigma (Q)\, v}
\label{timestep}
\end{equation}
As a first approximation for $\sigma (Q)$
we can take
\begin{equation}
\sigma (Q)\,\sim\, 4\,\pi  R_Q^2
\label{crosssection}
\end{equation}
where $R_Q$ is the radius of the Q-ball.
During a random walk from a small to a critical
Q-ball this radius varies by several orders
of magnitude, and also with different dependence
on the charge, according to relations
such as (\ref{pl1}), (\ref{pl2}).
 
However, the simple result for  the
probability (\ref{probability}) is
derived in the limit of a large number
of steps $N$ and turns out to be
finite and independent of $N$.
It is the same whether we use the
smallest or the largest time step, as long
as the number of steps is large enough,
and so it must be the same even if the
time step is varying.

\begin{figure}
         \centerline{\epsffile{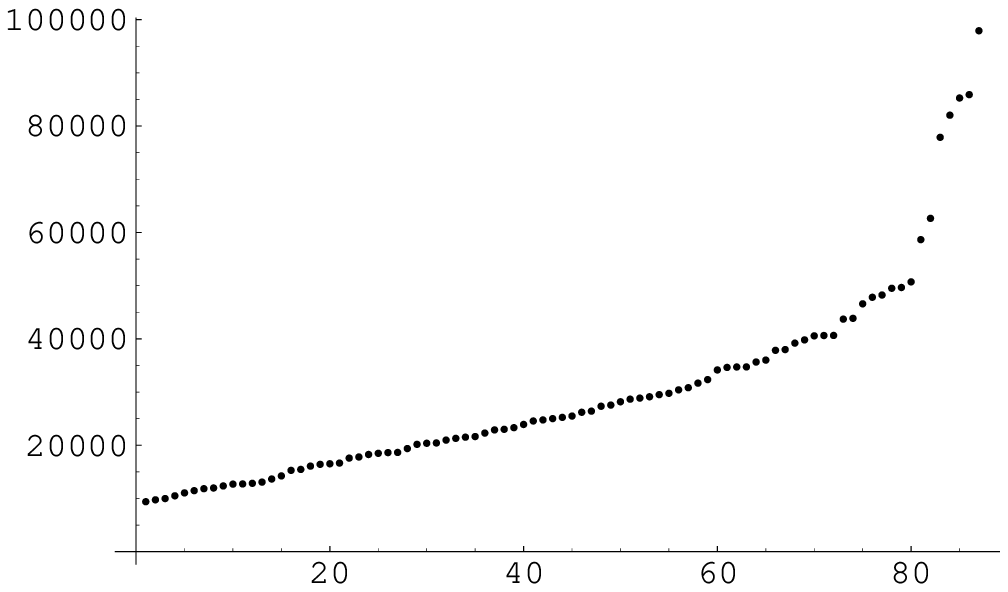}}
         \caption{In the vertical axis we plot the
                    number of steps of the random walk, and in the
                  horizontal axis the successful trials that cross
                                 the critical charge.}
         \label{randomwalkfig1}
         \end{figure}

\begin{figure}
       \centerline{\epsffile{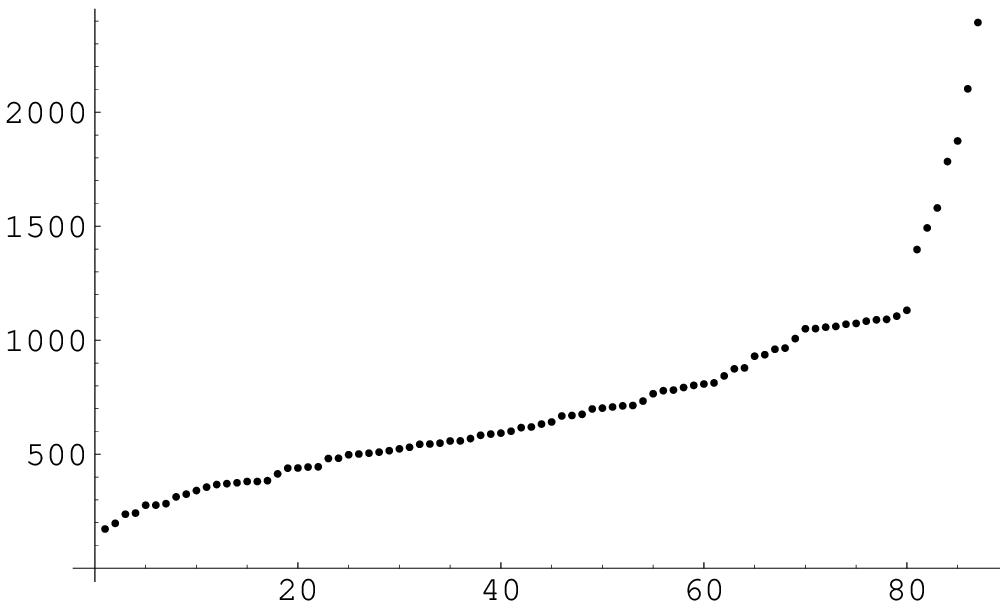}}
       \caption{In the vertical axis we plot
                    the duration time of the random walk in
                 units of $\phi^2 / T^3 $ and in the
             horizontal axis the successful trials.}
         \label{randomwalkfig2}
         \end{figure}

I  have performed  several numerical
trials of random walks with initial
charge $Q_0$ and critical charge  $Q_c$,
and time steps with various 
dependences on the charge
and indeed it turns out that for any
sample of random walks  there is
always a fraction of $Q_0/Q_c$ of them
that cross the  critical charge,
the dependence on time is very similar to 
the dependence on the number of steps, and
they are all completed within $Q_c^2$
steps. 
Some examples of these
random walks are shown in 
Fig.~\ref{randomwalkfig1} and
Fig.~\ref{randomwalkfig2}.

In these examples I used the numerical values
of $Q_0 =10$, $Q_c =300$, and for a sample
of 3000 trials of random walks, a fraction of $Q_0/Q_c$
where successful (crossed the critical charge).
They where all completed within $Q_c^2$ steps
and I show the results in Fig.~\ref{randomwalkfig1}.
I used a radius-charge dependence $R_Q \propto Q^{2/5}$,
for which the time step varies as $\Delta t \propto Q^{-4/5}$,
and I show the results for the duration of
these random walks in Fig.~\ref{randomwalkfig2}.

The assumption of a geometric cross section
as an estimate 
for $\sigma (Q)$ is justified by recent
results \cite{multa}, at least for large values of the 
charge. From our numerical trials of random walks
we see that indeed the successful random
walks spend most of the time in large values
of the charge, so we expect it to
be a good approximation.

Another approximation that was implicit in the 
random walk model was that the absorption 
of a positive (negative) charge leads instantaneously
to a Q-ball of charge $Q\pm 1$.
Actually, it leads to an excited Q-ball
of charge $Q\pm 1$ according to reactions like
\begin{equation}
(Q)\, +\, \phi\, \rightarrow\, (Q+1)\, +\, \Delta E(Q)\,.
\end{equation}
However, one can easily see that generally
$\Delta E(Q)$ is less than $M$, the elementary
mass of the particles,
and using (\ref{timestep}), (\ref{crosssection})
one can see that the thermal fluctuations coming
from collisions with the surrounding particles
of the heat bath during an elementary time step
of the random walk are of order $T$ which
is generally much greater than $M$, hence these
collisions bring rapidly an excited Q-ball
to its ground state.
 
\section{Nucleation of cosmological  phase transitions}
 
From the previous expressions for
the critical charge
and the critical radius,  Eqs.~(\ref{qc}), (\ref{rc}),
and using the relations for our general
finite temperature potential slightly
below the critical temperature,
Eqs.~(\ref{relations}), (\ref{surfenergy}), (\ref{fvenergy}),
we get
\begin{equation}
Q_c= \frac{c_3}{\lambda}
     \left(\frac{M}{\sqrt{\alpha}\,T}\right)^5
      \frac{1}{\eta^{5/2}}
\;\; , \;\;\;
R_c = c_4 \, \frac{\lambda^{1/2}}{\alpha}
      \,\frac{\phi}{T^2 \,\eta}
\label{qceta}
\end{equation}
with numerical constants
$c_3 =0.3$, $c_4 = 0.7$.
For the parameters of the potential
$M$, $A$, $\phi_{+}$, I will use their
values at the critical temperature,
assuming as will be shown later
that the transition happens at
a small value of the supercooling 
parameter.
For numerical results I will take
the coupling constants
$\lambda = 0.1$, $\alpha = 0.2$,
and the scale of the phase transition
at the temperature $T_c = 1TeV$.

I will assume a Friedman-Robertson-Walker (FRW), radiation-dominated
universe, with the time-temperature relation 
\begin{equation}
t=\frac{M_P}{T^2}
\label{time}
\end{equation}
where
\[
  M_P =\sqrt{\frac{45}{16\pi^3\,N}} \,\,M_{Planck}
\]
and $N=N_b+\frac{7}{8}N_f$
are the essentially massless degrees of freedom.
Thus, for temperature $T=T_c - \delta T$
slightly below the critical temperature
the supercooling  parameter is     
\begin{equation}
\eta = \frac{\delta T}{T}
     = \frac{\delta t}{2 t}
\end{equation}
 
I will estimate the value of the
supercooling  parameter at which there are
enough critical Q-balls produced,
at least one per Hubble volume,
in order to induce the
phase transition
\cite{GW}.

The critical charge is  infinite
at and  above the critical  temperature.
There are subcritical bubbles and
solitons  produced at these temperatures
but the probability for their survival
below the critical  temperature
is exponentially small, since this
evolution is equivalent to a random walk
with one boundary and for this process
we expect an initial charge $Q$ to  
disappear after $N\sim Q^2$
collisions.
 
Let $\tau (Q_c)$ be the time  needed
for the completion of the random  walk
for a  critical charge $Q_c$.
Then if
\begin{equation}
\tau (Q_c)\,\frac{dQ_c}{dt}\ll\,Q_c
\label{condition}
\end{equation}
the variation of the critical charge
during the random walk
 is irrelevant, 
and the number density of the
critical Q-balls produced at some time is
\begin{equation}
n(Q_c, t)=\int dQ\,dt'\,\frac{Q}{Q_c(t')}
               \Gamma(Q, t')
\label{ndensity}
\end{equation}
where $Q/Q_c$ is the probability
that a critical Q-ball with charge $Q_c$ will be nucleated
starting from a Q-ball with charge $Q$.
$\Gamma(Q, t')$ is the nucleation
rate of Q-balls at time $t'$,
and for a fast transition I will
approximate it with the nucleation rate $\Gamma(Q)$
at the critical temperature 
(\ref{nuclrate}).
 
From the previous results for the random walk
we can also get an estimate for $\tau(Q_c)$
\begin{equation}
\tau(Q_c) \simeq\, Q_c^2 \, \Delta t
\label{eq41}
\end{equation}
where,
for relativistic particles 
with density $n(\phi)$ in the heat bath,
and cross section for  absorption
$\sigma(Q)\simeq\,4\,\pi\,R_Q^2$,
the time step is
\begin{equation}
\Delta t \simeq\, \frac{1}{n(\phi)\sigma(Q)}
         \simeq\, \frac{1}{ T^3 \, R_Q^2}
\label{eq42}    
\end{equation}                       
For an order of magnitude estimate  of $\tau(Q_c)$
I consider an average charge $Q=Q_c /2$,
with the radius-charge dependence
given by (\ref{pl2}), since it
is near the phase transition but far from the critical charge:
\begin{equation}
R_Q^2 \simeq\, \frac{1}{\lambda^{1/5}}
        \frac{Q^{4/5}}{\phi^2}
\label{eq43}
\end{equation}
  
Also we have
\begin{equation}
  \frac{dQ_c}{dt} =
   \frac{dQ_c}{d\eta}\,\frac{d\eta}{dt}
\label{eq44}
\end{equation}
where, from (\ref{time})
\begin{equation}
\frac{d\eta}{dt} = \frac{T^2}{2\, M_P}
\label{eq45}
\end{equation}
Using (\ref{qceta}), (\ref{eq41}),
(\ref{eq42}), (\ref{eq43}), (\ref{eq44}), (\ref{eq45}),
the condition for the validity of our 
approximation (\ref{condition}) becomes:
\[
\tau (Q_c)\,\frac{dQ_c}{dt}\ll\,Q_c  \,\Leftrightarrow
\]
\[
\Leftrightarrow\,
Q_c^2 \frac{1}{ T^3}\,
      \frac{\lambda^{1/5}\,\phi^2}{(Q_c /2)^{4/5}}\,
      \frac{T^2}{2\,M_P}\,
       \frac{5\,Q_c}{2\,\eta}\,
                          \ll\,Q_c  \,\Leftrightarrow
\]
\begin{equation}
\Leftrightarrow\,
\eta^4 \gg  \,\frac{1}{\lambda}
        \left(\frac{M}{\sqrt{\alpha}T}\right)^6 
           \frac{\phi^2}{T\,M_P}  
\label{fcondition}
\end{equation}
With $M_P  \sim\,10^{19}\, GeV$ and for
a  phase transition at a scale
$T\sim\,\phi\,\sim\,1\,TeV$, this condition
can be satisfied for a small value of the
supercooling parameter $\eta$.
If this condition holds, then
\begin{eqnarray}
n(Q_c, t)&=&\int dQ\,dt'\,\frac{Q}{Q_c(t')}
               \Gamma(Q, t') \,= \nonumber\\
         &=& \int_{0}^{\infty} Q\,\Gamma(Q)\,dQ\,\,
             \int_0^{\eta} \frac{\lambda\,(\eta ')^{5/2}}{c_3}\,
                           \frac{2\,M_P}{T^2}\,
\left(\frac{\sqrt{\alpha}T}{M}\right)^5
                                               d\eta '
\label{eq440}
\end{eqnarray}
Here I took the lower limit of  the
time integral at the critical 
temperature, assuming that
the condition (\ref{fcondition})
is quickly satisfied.
 
From the previous expression (\ref{neweq})
for $\Gamma(Q)$ we have
\begin{equation}
\int_{0}^{\infty} Q\,\Gamma(Q)\,dQ\, =
a \, \frac{T^{21/2}}{(\phi M)^{13/4}} \,
  e^{\textstyle -c\,\frac{\phi\,M}{T^2}\,Q_{min}}
\label{eq441}
\end{equation}
so we get
\begin{equation}
n(Q_c, t)=
    a \, \frac{T^{17/2}}
                    {(\phi M)^{13/4}} \, 
              e^{\textstyle -c\,\frac{\phi\,M}{T^2}\,Q_{min}}
\,\,
\left(\frac{\sqrt{\alpha} T}{M}\right)^5
\lambda \, M_P \,
\eta^{7/2}
\label{eq442}
\end{equation}
where, $\phi=\phi_+$ is the asymmetric minimum
at the critical temperature, and $M=M(T_c)$.
Thus, the condition  for a large enough
density 
of Q-balls to be produced inside
a Hubble volume in order 
to mediate the phase transition:
\begin{equation}
n(Q_c, t) \stackrel{>}{\sim}
\left(\frac{T^2}{M_P}\right)^3
\label{eq12}
\end{equation}
gives
\begin{equation}
a \,\,e^{-c\frac{\phi\,M}{T^2}Q_{min}}
\,\,\eta^{7/2}
\stackrel{>}{\sim}
\left(\frac{ T}{M_P}\right)^4
\,\,\frac{1}{\lambda}\,\,
\left(\frac{\phi\,M}{T^2}\right)^{13/4}\,\,
 \left(\frac{M}{\sqrt{\alpha} T}\right)^{13/4}
\label{sc}
\end{equation}
with the numerical constant
$c=1.42$  and  
$a$ a
numerical factor of order one.
 
If $Q_{min}$ is small enough for this
condition to be satisfied,
then the transition proceeds at the
corresponding value of $\eta$.
Again, because of the factors of $M_P$,
for a late time phase transition
at the scale of $1TeV$, this condition
is satisfied for a very small value
of the supercooling parameter, if the
minimum charge is small enough.
 
The nucleation of a sufficient number
of critical Q-balls depends sensitively
on the minimum charge.We see that
for large enough $Q_{min}$ (greater than 22 in our example)
the phase
transition cannot proceed through Q-ball
nucleation, at least not in the range of our
approximations, something which is consistent
with our expectations and with previous
results \cite{Ellis}.
 
As $Q_{min}$ becomes smaller, there is a
sufficient number of critical Q-balls
produced at small  values of the 
supercooling parameter.
 
For example, for $Q_{min}=20$,
we see that there are enough critical Q-balls
produced per Hubble volume to nucleate the 
phase transition at values of the 
supercooling parameter $\eta = \frac{\delta T}{T}  \sim 10^{-2}$.
 
At these values of $\eta$, if we consider
the case of ordinary tunneling 
we find that the
tunneling rate is much smaller than the
expansion rate of the Universe, and the
transition does not happen until 
much later times and larger values
of the supercooling parameter.
Indeed, 
in our range of values of the supercooling
parameter 
$\eta \sim 10^{-2}$,
the nucleation rate
of ordinary critical bubbles at
high temperature is 
calculated in the thin-wall approximation \cite{Linde}  
$\Gamma \sim T^4 \, e^{-B}$,
where
\begin{equation}
B=\frac{S_1^3}{3\varepsilon^2 T}\sim
    \frac{\sqrt{\alpha}}{\lambda}\,
    \frac{1}{\eta^2}
\end{equation}
and we see that, for $\eta \stackrel{<}{\sim} 10^{-1}$,
and for our values of parameters $\alpha$ and $\lambda$,
the result for the ordinary tunneling rate
is indeed exponentially smaller
than the Hubble expansion rate.

For a better description of the phase transition
one needs to calculate quantities like $p(t)$, the 
fraction of the coordinate volume of the universe
that remains in the symmetric phase $\phi = 0$ at
time $t$ after the beginning of the phase transition, 
which in our case may start right after the time
$t_c$ when the temperature has dropped to the
critical temperature $T_c$.
The usual formula given in \cite{guthtye}
can be easily generalized to take into account
the initial critical soliton radius, and the    
velocity of the soliton expansion:
\begin{equation}
\ln p(t) = \, -\, \int^t_{t_c} dt_1 \Gamma(Q_c , t_1)
R^3 (t_1)  \frac{4\pi}{3}
\left(\frac{R_c(t_1)}{R(t_1)} \, + \,
\int^t_{t_1} dt_2 \, \frac{v(t_2)}{R(t_2)}
\right)^3
\end{equation}
Here, $R$ is the scale factor of the Robertson-Walker
metric, $R_c$ is the critical soliton radius,
and $v$ is the velocity of the critical soliton 
expansion.
The nucleation rate
of critical solitons $\Gamma(Q_c , t_1)$
and their initial radius $R_c(t_1)$ are
calculated from     
(\ref{ccc}), (\ref{qceta}).
Not much is known about the velocity of
expansion of critical solitons.
We do not expect it to be close to the speed
of light, since we 
are very close to the critical
temperature and the amount of supercooling 
is very small.
However, if we assume that the expansion
is similar to the expansion of a true vacuum
bubble in the case of an ordinary first-order
phase transition, then we can use a simple
model to approximate it by balancing  
the pressure inside and outside the bubble
\cite{Linde}    
\begin{equation}
  v\, = \,c_5\, \frac{\varepsilon}{T^4} \,
    =\, c_5 \frac{\alpha \phi^2}{T^2} \, \eta
 \end{equation}
where $c_5 = 30/\pi^2$.

Substituting from the previous results, and
keeping the first-order correction coming 
from the initial soliton radius,
we get
\begin{eqnarray} 
\ln p(\eta) = \, - &a& 
e^{-c\frac{\phi\,M}{T^2}Q_{min}}
 \left( \frac{T^2}{\phi\,M}\right)^{5/4} 
\left\{  \lambda \alpha^{11/2} \, 
       \left(\frac{M_P}{T}\right)^4
       \frac{T^3\,\phi^4}{M^7} \right. 
       \,\eta^{17/2}   \nonumber \\       
       \,&+& \,
\left.       \lambda^{3/2} \alpha^{7/2}
        \left(\frac{M_P}{T}\right)^3
        \frac{T^4\,\phi^3}{M^7}
       \,\eta^{11/2}
          \right\}
\label{final}            
\end{eqnarray}   
Here, the first term in the brackets would be
the final result if we had neglected the
initial soliton radius. The second term is
the first-order correction that comes from
the critical soliton size.

This result confirms our previous estimates
for the nucleation of the phase transition.
The mechanism is similar to inhomogeneous
nucleation or nucleation in the presence
of impurities \cite{various}.
The corrections coming from the critical soliton size
are negligible for generic values of the couplings
in our model, but for exceptional values
of the couplings, or maybe different
models, they would have to be taken into
account.

\section{Conclusion}   

The final results (\ref{eq442}), (\ref{sc}), (\ref{final})   
 depend quite sensitively      
on the minimum soliton charge allowed for
stability. Thus the relaxed minimum bound
on the charge described in Section 2
is important. As explained before, the minimum
charge that the solitons have to exceed
for reasons of stability is a model dependent
parameter, and our results for the occurence
or not of the phase transition
can also be viewed as conditions on
this charge in order for
the transition to proceed
via Q-ball nucleation.
 
I used a simplified model for the soliton
production based on charge fluctuations,
simulating the absorption of
positive or negative charges by a random walk and
ignoring other processes like soliton evaporation
that may be included in more complete models.
I assumed that there is no asymmetry between
the Q-ball interaction with positive and negative
charges, and also that the absorption
of a positive (negative) charge by the Q-ball
turns it instantaneously into a soliton
of charge $Q\pm 1$. I also assumed a 
geometrical cross-section for absorption of
elementary particles by the Q-ball, which should
be a valid approximation for macroscopic Q-balls
of large charge. For small charge this may break
down but since the Q-balls spent
most of the time of the random walk in
large values of the charge this should be
a valid approximation.
 
Although these approximations are made to
simplify the analysis,       
I do not expect the final
results  for the minimum soliton charge
and the  contribution
to  the phase transition  
 to change
significantly if appropriate corrections
are included, since these modifications
may only  affect the pre-exponential
factors in (\ref{eq442}). 
 
The most important correction
is probably concerned with 
the model of soliton production via
fluctuations of true vacuum bubbles. 
This is not 
as well-tested as other mechanisms of
soliton production like, for example,
the Kibble mechanism \cite{Kibble},
but there are indications that the
subcritical bubble formalism is correct
\cite{Kolb1}.
  
\medskip
\medskip
\leftline{\large \bf Acknowledgments}
\smallskip
\noindent 
I would like to thank Profs. E.~J.~Weinberg
and V.~P.~Nair for many useful comments.
This work was supported in part by
a PSC-CUNY award.

 \newpage
  
         \bigskip \bigskip

         \bigskip \bigskip

\end{document}